\documentclass[a4paper]{article}
\usepackage[]{graphicx}\usepackage[]{color}
\makeatletter
\def\maxwidth{ %
  \ifdim\Gin@nat@width>\linewidth
    \linewidth
  \else
    \Gin@nat@width
  \fi
}
\makeatother

\definecolor{fgcolor}{rgb}{0.345, 0.345, 0.345}

\usepackage{framed}
\makeatletter
 {\par\unskip\endMakeFramed%
 \at@end@of@kframe}
\makeatother

\definecolor{shadecolor}{rgb}{.97, .97, .97}
\definecolor{messagecolor}{rgb}{0, 0, 0}
\definecolor{warningcolor}{rgb}{1, 0, 1}
\definecolor{errorcolor}{rgb}{1, 0, 0}
\newenvironment{knitrout}{}{} % an empty environment to be redefined in TeX

\usepackage{alltt} % man for manuscript format, jou for journal format, doc for standard LaTeX document format
\usepackage{natbib} 
\usepackage[american]{babel}
\usepackage[utf8]{inputenc}
\usepackage{csquotes}

\usepackage{setspace}

\usepackage[outdir=./]{epstopdf}

\usepackage{amsmath,amssymb,amsfonts}

\usepackage{url}   % this allows us to cite URLs in the text
\usepackage{graphicx}   % allows for graphic to float when doing jou or doc style
\usepackage{verbatim}   % allows us to use \begin{comment} environment
\usepackage{caption}
\usepackage{pdflscape}

\usepackage{fancyvrb}

\usepackage{newfloat}
\DeclareFloatingEnvironment[
]{listing}

\title{Statistical methods for linguistic research: Foundational Ideas - Part I}

\author{Shravan Vasishth and Bruno Nicenboim}

\begin{document}

\maketitle

\bibliographystyle{apalike}

\begin{abstract}
We present the fundamental ideas underlying statistical hypothesis testing using the frequentist framework. We begin with a simple example that builds up the one-sample t-test from the beginning, explaining important concepts such as the sampling distribution of the sample mean, and the iid assumption. Then we examine the p-value in detail, and discuss several important misconceptions about what a p-value does and does not tell us. This leads to a discussion of Type I, II error and power, and Type S and M error. An important conclusion from this discussion is that one should aim to carry out appropriately powered studies.  Next, we discuss two common issues we have encountered in psycholinguistics and linguistics: running experiments until significance is reached; and the ``garden-of-forking-paths'' problem discussed by Gelman and others, whereby the researcher attempts to find statistical significance by analyzing the data in different ways. The best way to use frequentist methods is to run appropriately powered studies, check model assumptions, clearly separate exploratory data analysis from confirmatory hypothesis testing, and always attempt to replicate results.
\end{abstract}

\section{Introduction}

Psycholinguistics has a long tradition of using experimental methods, but in recent years, linguists working in areas such as syntax, semantics, and pragmatics have also started to embrace empirical methods (see \citet{arunachalam2013experimental} for a review of the more commonly used methods).   As a consequence, basic familiarity with experiment design is becoming a core requirement for doing linguistics.  However, just knowing how to carry out an experiment is not enough;  a good understanding of statistical theory and inference is also necessary. In this article, we present the most important issues that researchers need to be aware of when carrying out statistical inference using frequentist methods (as opposed to Bayesian approaches, see \citet{NicenboimVasishthStatMeth}).
We focus on frequentist methods because
the statistical tools of choice in psycholinguistics and linguistics are usually frequentist ones; examples are the t-test, analysis of variance (ANOVA), and linear mixed models. Given software such as R \citep{R}, it is extremely easy to obtain statistics such as t- of F-values, and the corresponding p-values. However, it is equally easy to misunderstand what these mean; in particular, a misinterpretation of the p-value often leads researchers to draw conclusions from their data that are not supported by the underlying statistical theory. 
We start the paper illustrating the meaning of t- and p-value and discussing some common misconceptions by means of the one-sample t-test or
(equivalently) the paired t-test. We use this test as an example because of its relative simplicity and because it happens to be a very frequently used one in linguistics and psycholinguistics. For ANOVAs and linear mixed models, the situation is more complex, but the same logic and issues described below also apply. We then show the importance of power, and Type I, II, S, and M errors using simulations based on linear mixed models. Two further important topics discussed are: the problems involved in running participants till significance is reached, and the issues involved in experiments with multiple measures, multiple regions of interest, and too many degrees of freedom in analysis.

\section{A simple example: a two-condition repeated measures design}

Consider the case of a two-condition repeated measures self-paced reading
\citep{jcw82} experiment, e.g., subject versus object relative clauses, where
the dependent measure is reading time in milliseconds; assume that reading time is
measured at a particular region of interest in the relative clause sentences.
Suppose the experiment has $n$ randomly sampled participants, each of whom read 
multiple
instances of subject and object relative clauses in a counterbalanced Latin
square design \citep{arunachalam2013experimental}. A typical approach taken is to calculate the mean of each
participant's reading time for each relative clause type by aggregating over all the items that the participant saw. 
To make this example concrete, consider the simplified situation in Table~\ref{example1} where a participant, labeled 1, sees three items for each condition; normally, of course, each participant will be shown many more items. The condition labels in Table~\ref{example1} refer to subject relatives (condition a) and object relatives (condition b).
If we average the three data points from the participant for condition a and for condition b, we obtain the aggregated data shown in Table~\ref{example1a}.

\begin{table}[!htbp]
\begin{center}
\begin{tabular}{cccc}
participant id & item id & condition & reading time\\
1          &  1      &    a      &   500\\
1          &  2      &    a      &   600\\
1          &  3      &    a      &   700\\
1          &  4      &    b      &   450\\
1          &  5      &    b      &   550\\
1          &  6      &    b      &   650\\
\end{tabular}
\end{center}
\caption{Hypothetical unaggregated reading time data from a two-condition experiment for one participant who saw six items, three from each condition.}\label{example1}
\end{table}%

\begin{table}[!htbp]
\begin{center}
\begin{tabular}{cccc}
participant id & condition & reading time\\
1          & a      &   600\\
1          & b      &   550\\
\end{tabular}
\end{center}
\caption{The result of aggregating the data over items for participant 1 in Table~1.}\label{example1a}
\end{table}%

\noindent
This procedure, applied to each of the participants in an experiment, results in two data points from each participant, one for each condition. This is called a by-participants (or by-subjects) analysis.
One can analogously do a by-items analysis by aggregating over participants. 

After this aggregation procedure, we have $n$ data points for subject relatives and $n$ for object relatives. The data are paired in the sense that for each participant we have an estimate of their reading time for subject relatives and for object relatives.  In the hypothetical example above, for participant 1, we have a mean subject relative reading time of $600$ ms and an object relative reading time of $550$ ms. If, for each participant, we take the difference in object vs subject relative reading time (for participant 1 this would be $-50$ ms), we have a vector of $n$ values, $x_1,\dots,x_n$ that are assumed to be \textit{mutually independent}, and represent the difference in OR vs SR reading times for each participant. 
Another assumption here is that these observed differences between RC types $x_1,\dots,x_n$ are generated from a normal distribution with some unknown mean $\mu$ and standard deviation $\sigma$. Since each of the data points is assumed to come from the same distribution, we say that they are \textit{identically distributed}. The independence assumption mentioned above and the identical-distribution assumption are often abbreviated as iid---independent and identically distributed. The statistical test depends on the iid assumption, and the assumption that a simple random sample of participants has been taken. 
For example, if we were to do a t-test on the unaggregated data, we would violate the independence assumption and the result of the t-test would be invalid. When distributional assumptions (such as the normality assumption of residuals in linear mixed models, see \citet{SorensenVasishthTutorial} for more discussion) are not met, the parametric bootstrap \citep{efron1994introduction} is an option worth considering. The bootstrap can also be used for linear mixed models; for an example, see Appendix D of \citet{bostonhalevasishthklieglLCP09}. 

\begin{figure}[!htbp]
\begin{knitrout}
\definecolor{shadecolor}{rgb}{0.969, 0.969, 0.969}\color{fgcolor}

{\centering \includegraphics[width=\maxwidth]{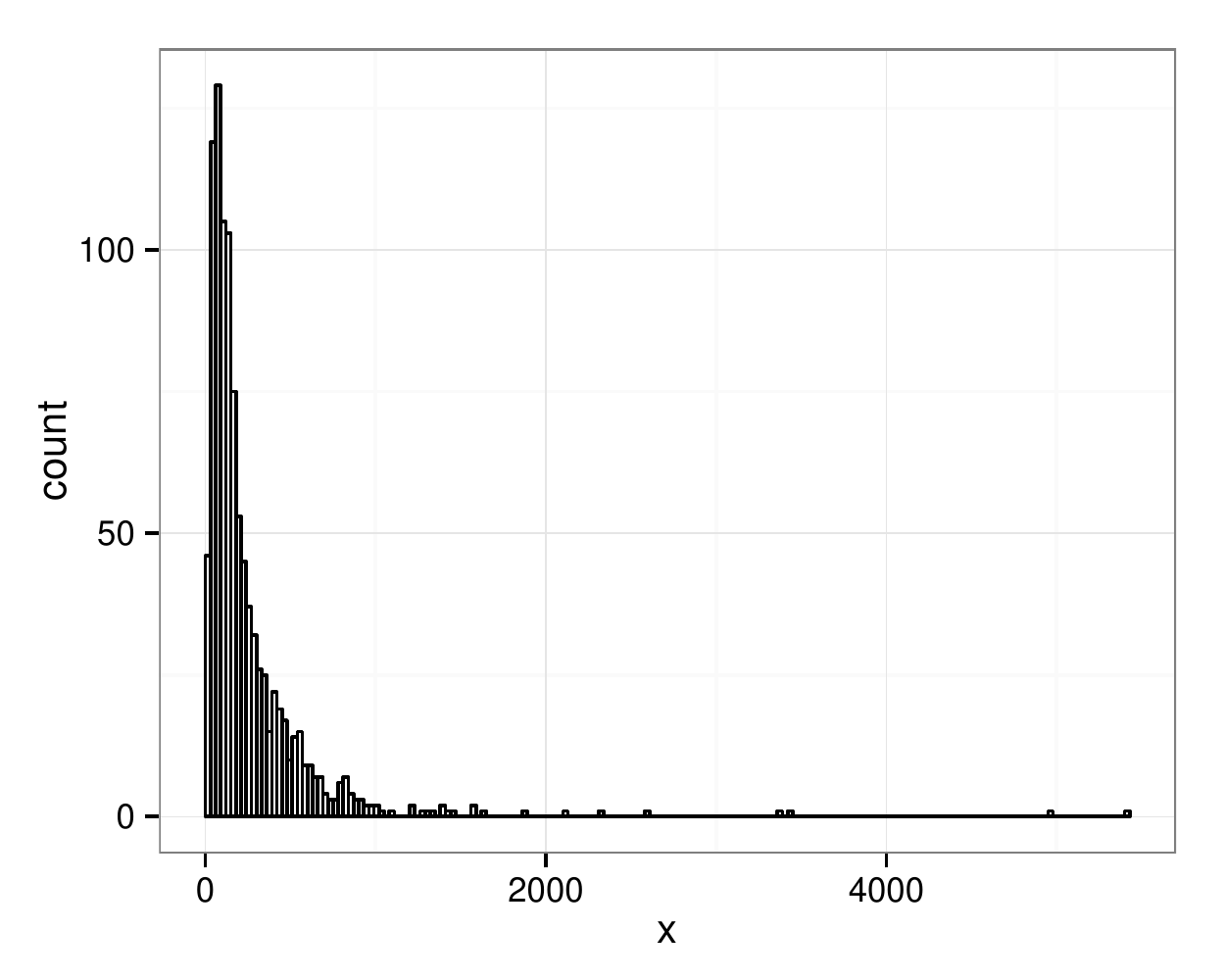} 

}

\end{knitrout}
\caption{Histogram of simulated reading time data.}\label{fig:histrt}
\end{figure}

Returning to the t-test under consideration, we begin by generating some fake data; all the R code used in this paper is available from the first author's home page. Let us simulate $n$ data points representing differences in reading times, as an illustration. Since reading time data typically have a log-normal distribution, we will simulate $n=1000$ draws from a log-normal distribution with
mean $2$ and standard deviation $1$ (Figure~\ref{fig:histrt}).

Given such a sample, the one-sample t-test (or, equivalently the paired t-test) works as follows. 
We first compute the sample mean and the sample standard deviation $\bar{x}$ and $\hat\sigma$; these are estimates of the unknown parameters, $\mu$ and $\sigma$, of the underlying distribution that is assumed to have generated the data.
It is important to note the distinction between the sample mean $\bar{x}$ and the unknown true point value $\mu$. Researchers often report their sample means by labeling it $\mu$ or (in the case of linear mixed models) $\beta$; but this is incorrect. The true parameter $\mu$ or $\beta$ is unknown; we are reporting an estimate of this unknown value. 

Statistical inference relies on an important property of the \textit{sampling distribution of the sample means} under repeated sampling: For a large enough sample size $n$, the distribution of the sample means under repeated sampling will be normal with mean $\mu$ and standard deviation $\sigma/\sqrt{n}$; this is assuming that
the underlying distribution that generated the data has a mean and standard deviation. The preceding sentence is an informal statement of the \textit{central limit theorem}.
The standard deviation of the sampling distribution of the sample means is also called the \textit{standard error} (SE), and can be estimated from the data by computing $\hat\sigma/\sqrt{n}$. 

Statistical inference in the frequentist paradigm begins by positing a null hypothesis distribution, which is a statement about what the true sampling distribution of the sample means looks like. 
In our example, our null hypothesis is that the difference in means between the two RC types is $0$. We will follow standard practice in writing this null hypothesis as $H_0: \mu=0$; $\mu$ represents the true, unknown difference in means between the two RC types.  
Next, we use the fact that the transformation $T=(\bar{x}-\mu)/SE$ has a t-distribution with $n-1$ degrees of freedom, where $n$ is the sample size (for large $n$, the t-distribution approximates the normal distribution with mean 0 and variance 1).
Since we have hypothesized $\mu$ to be 0, and since we
have estimated $\bar{x}$ and $\hat\sigma$ from the data, we can compute the
observed t-value $t=(\bar{x}-0)/SE$. This observed t-value is the distance between the sample mean and the hypothesized mean, in SE units; this is easy to see if we rearrange the equation as follows: $t\times SE=\bar{x}-0$. Intuitively, if the t-value---the distance in SE units between the sample means and the hypothesized mean---is large, we feel entitled to reject the null hypothesis. It is traditional to 
compute the p-value associated with the observed t-value; this is the
probability of observing a t-value at least as extreme as the one we observed, conditional on the assumption that the null hypothesis is true. 
It is also traditional to reject the null hypothesis if this conditional probability falls below $0.05$.

\begin{figure}[!htbp]
\begin{knitrout}
\definecolor{shadecolor}{rgb}{0.969, 0.969, 0.969}\color{fgcolor}

{\centering \includegraphics[width=\maxwidth]{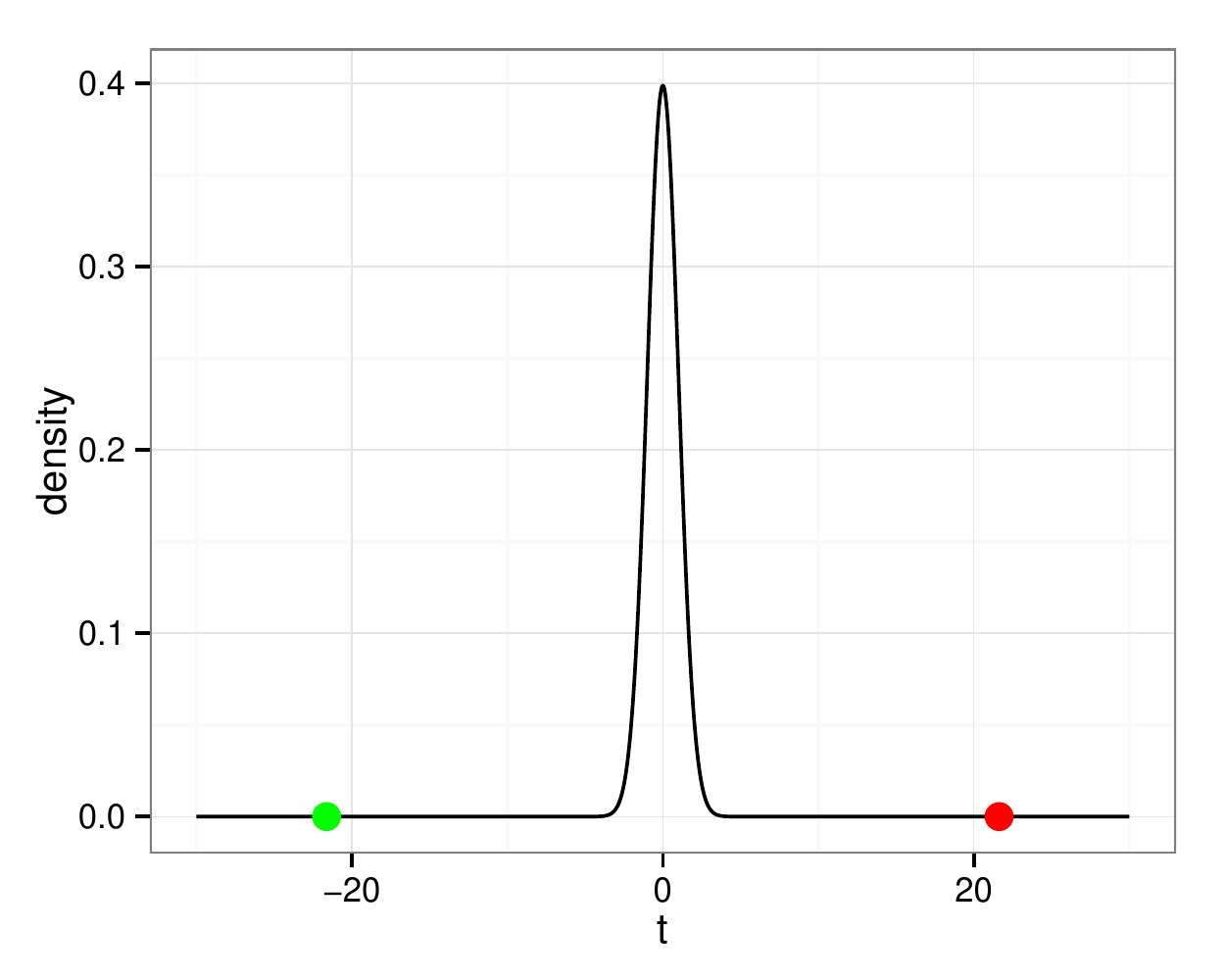} 

}

\end{knitrout}
\caption{The null hypothesis distribution and the observed t-value.}\label{fig:obst}
\end{figure}

\noindent
In the above example with simulated data, our sample mean is
257.46, and the standard deviation is
376.47, leading to an observed t-value of
21.63. Can we reject the null here following our rule above, which is to reject the null if the p-value is below $0.05$?
We can start by visualizing the null hypothesis distribution and the observed t-value; see Figure~\ref{fig:obst}. The p-value is the area under the curve to the right of the red dot (our observed t-value), plus the same area under the curve to the left of the green dot. The red and green dots mark the observed t-value 21.63 and its negation \ensuremath{-21.63}.
We consider both sides of the distribution because we want to know the probability of seeing the absolute t-value (regardless of sign), or a value more extreme. This is called the two-sided t-test.

The p-value in our example is going to be some very small number because the observed t-value is far out in the tails of the null hypothesis distribution; the probability mass is going to be very small (near 0). So yes, in this simulated example, we would reject the null hypothesis. That is the t-test in a nutshell, and it will serve as the basis for further discussion about statistical inference.

\section{Four misconceptions about what a p-value tells us}

It is very easy to overlook the fact that the p-value is a conditional probability. Neglecting to attend to this detail has led to quite a few misconceptions about what a p-value tells us. 
Here are some claims about the p-value that are incorrectly believed to be true. Note that the same misconceptions hold if one considers, instead of the p-value, the absolute value of the observed t-value, as is commonly done in connection with linear mixed models.

% Perhaps the most important thing to understand in the t-test is the meaning of the p-value. We discuss this point below, pointing out four misconceptions that we frequently encounter in psycholinguistics.

\subsection{Misconception 1:
The smaller the p-value, the greater the confidence in the \textit{specific} alternative hypothesis we are interested in verifying}

In fact, the smaller the p-value, the greater the confidence that the null
hypothesis is false. It doesn't tell us \textit{which} of the infinity of possible alternative $\mu$ is now true, only that the null hypothesis, that $\mu=0$, is false. 
Rejecting the null doesn't give us any statistical evidence for the
\textit{specific} effect our theory predicts, it just gives us evidence against a very specific hypothesis that $\mu=0$, and allows all other values of $\mu$ to be plausible, even ones we would not be happy to see as the outcome of an experiment. We want to know  P($H_1$ $\mid$ data), i.e., how probable is it that our theory is correct (the specific alternative $H_1$) given the data that we have, but frequentist statistics tells us P(data $\mid$ $H_0$), i.e., the probability of the data (more accurately, the probability of seeing a test statistic as extreme or more extreme than the one observed) given the null hypothesis. Importantly, we cannot infer one conditional
probability just by knowing its inverse. \citet{dienes2011bayesian}
illustrates this with a very graphic example: The probability of dying given that a shark has bitten one's head clean off, P(dead $\mid$ head bitten clean off by shark), is one. But most people die of other causes; given that one is dead, the probability that a shark has bitten one's head clean off, P(head bitten off by shark $\mid$ dead), is close to zero. In summary, the p-value answers \textit{a} question, but it doesn't answer the question we are interested in.
To answer the question we are actually interested in, namely whether the effect is positive or negative with a certain magnitude, we make an indirect inference by looking at the observed mean and draw the conclusion that our theory is supported or not by looking at the sign of the observed mean. The p-value does not provide any evidence that our theory is supported; it only gives us evidence against the null.
The most informative piece of information we have about our specific hypothesis is actually the sample mean and the uncertainty associated with out estimate of this sample mean: the standard error.

\subsection{Misconception 2: A p-value greater than 0.05 tells me that the
null hypothesis is true}

This is perhaps the commonest mistake seen in linguistics and
psycholinguistics.  Researchers in linguistics and psycholinguistics (and also psychology) routinely
make the strong claim that \textit{There is no effect of factor X on dependent
variable Y}, based on their getting a p-value larger than 0.05. This claim can only be made when power is high, as discussed below.

This misconception arises because researchers do not consider the fact that the p-value is a conditional probability: the probability of getting a statistic as extreme or more extreme as the one we got, conditional on the null hypothesis being true. To conclude that the null is true when $p>0.05$ is like arguing that, if the probability of the streets being wet given that it has just rained is higher than $0.05$, then we can conclude that it has just rained.

Unless one has sufficient power (see below), the best one can say when we get
a p-value larger than $0.05$ is that ``we failed to find an effect''. Instead, in linguistics and psycholinguistics it is routine to make the much stronger, and statistically invalid, claim that ``there is no effect''.

\subsection{Misconception 3: Two nested comparisons allow us to draw conclusions about interactions}

\begin{figure}[!htbp]
\begin{knitrout}
\definecolor{shadecolor}{rgb}{0.969, 0.969, 0.969}\color{fgcolor}

{\centering \includegraphics[width=\maxwidth]{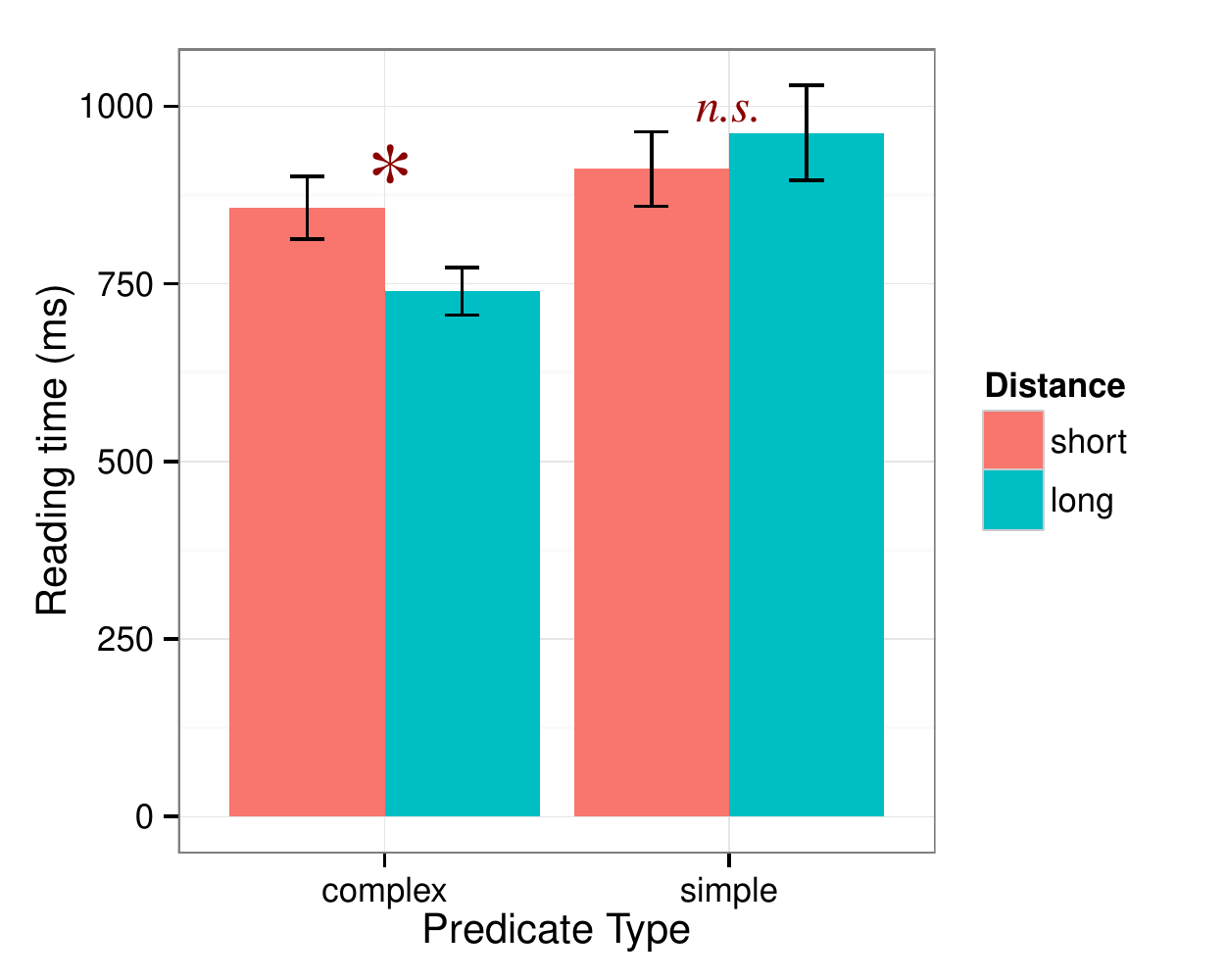} 

}

\end{knitrout}
\caption{Example showing a typical situation where one comparison is statistically significant, but the other is not. From this, we cannot conclude that the interaction is statistically significant.}\label{fig:misconception3}
\end{figure}

To illustrate the issue here, consider the results  reported in Experiment 2 of \citet{HusainEtAl2014}; we use their data because it is publicly available. In the published paper we had used a Bayesian linear mixed model to report the results, but here we will do the same analysis using a series of one-sample t-tests discussed above. This experiment was a $2\times 2$ factorial design, with one factor predicate type (complex vs simple) and the other factor distance between the verb and an argument noun (long vs short).
We can look at the effect of distance within complex predicates and simple predicates separately. These comparisons can be done easily following the procedure of the one-sample t-test we discussed in the beginning of this paper. We convert reading times to the log scale to do the test (for reasons discussed later in this paper).
When we do these tests, 
we find a statistically significant effect of distance in complex predicates (t(59)=\ensuremath{-2.51}, p-value=0.02), but we don't get a significant effect of distance in simple predicates (t(59)=0.52, p-value=0.61). This is shown graphically in Figure~\ref{fig:misconception3}.
Can we now conclude that the interaction between the two factors exists?
No. First of all, one can check with a t-test whether the interaction is statistically significant.  In the first test, the null hypothesis is that the true difference, $\delta_1$, between the (unknown) means of the long vs short conditions in the complex predicate case is 0; in the second test, the null hypothesis is that the true difference, $\delta_2$, between the (unknown) means of the long vs short conditions in the simple predicate case is 0. In the interaction, the null hypothesis is that the difference between these two differences, $\delta_1-\delta_2=\delta$ is 0.
When we do the t-test with this null hypothesis (see accompanying code for details), 
we find that we cannot reject the null hypothesis that $\delta=0$:  (t(59)=\ensuremath{-1.68}, p-value=0.1).

The first comparison (the effect of distance in complex predicates) yields a difference of sample means \ensuremath{-0.13}, and the standard error of this difference is 0.05. 
The second comparison (the effect of distance in simple predicates) yields a difference of sample means 0.03, and the standard error of this difference is 0.06. From the first two comparisons, we cannot conclude that the interaction will necessarily be significant; indeed,  in our example, the difference between these differences in means is 
\ensuremath{-0.16}, with standard error 0.09. Thus, one must always check whether the interaction is significant.
This is a real issue in psychology and linguistics and has serious consequences for theory development; many papers have misleading conclusions that follow from this error.
As evidence, \citet{nieuwenhuis2011erroneous} present a survey of published articles showing that approximately 50\% of them (79 articles) draw this incorrect inference.

\subsection{Misconception 4: If $p<0.05$, we have
found out that the alternative is in fact true}

This misconception is perhaps encouraged by the language we use to describe our success in getting a p-value; the effect is ``reliable'' or even ``real''. The word ``significant'' also contributes to giving the feeling that something of importance has been found. It doesn't help that textbooks and articles explaining the p-value often state that the p-value tells us whether ``the effect is due to chance''. If this were literally what the p-value meant, it would be reasonable  to conclude that if the p-value is low, then the effect is \textit{not} due to chance. But this characterization of the p-value is not correct. The phrase ``due to chance'' is more accurately expanded to ``due to chance under the null hypothesis''. Stated correctly in this way, if we get a very low p-value, we can only say that, \textit{assuming that the null is true}, the probability of observing the t-value (or some value more extreme) is very low; it is on the basis of this low probability that we reject the null. No absolute certainty is afforded by the p-value, no matter how low it is.

In fact, no matter how low our p-value, we will always have a 0.05
probability of having mistakenly rejected the null when the null is in fact true. Thus, a p-value (regardless of whether it is low or on the border of 0.05) alone should not convince us that an effect is ``real''. We discuss this point further in the next section.

\section{Type I, II error, power, and Type S, M error}

The logic of frequentist methods is inextricably linked with hypothetical
repeated sampling. If we were to repeatedly run an experiment, we would  essentially get a different sample mean in every repeated sample; in some of these samples,  we will reject the null, and in
some other samples we will fail to reject the null. Under such 
repeated sampling---which we almost never have the luxury of doing in real life, incidentally---we can define the probability of incorrectly rejecting the null (when it's actually true); this is called \textit{Type I} error (also called the $\alpha$ value) and is typically set at $0.05$ by the researcher. Type I error is conventionally fixed at 0.05 before we run an experiment. Note that it has nothing to do with the p-value: the p-value is computed based on the data you have at hand, and will vary depending on your data, whereas Type I error is a fixed rate of incorrect rejections of the null under repeated sampling.

Under repeated sampling, the probability of incorrectly
failing to reject the null hypothesis when it is false with some specific value is called \textit{Type II} error. The quantity (1-Type II error) is called \textit{power}, and is the probability of correctly rejecting the null.\footnote{Note that all our definitions here are with respect to the null hypothesis---it is a mistake to think that Type II error is the probability of failing to accept the alternative hypothesis when it is true. We can only ever reject or not reject the null; our hypothesis test is always with reference to the null.} 
Table~\ref{tab:type12power} shows the four possible states when we consider the two possible states of the world (null true or false) and the binary decision we can take based on the statistical test (reject the null or not).

\begin{table}[!htbp]
\centering
\begin{tabular}{ccc}
\hline
Reality: & $H_0$ TRUE & $H_0$ FALSE \\
\hline
Decision: `reject': & $\alpha$ & $1~-~\beta$ \\
                                     & \textbf{Type I error}                         & \textbf{Power} \\                                      
                                     & & \\
\hline
Decision: `fail to reject': & $1 - \alpha$ & $\beta$ \\                                    &                                 & \textbf{Type II error}\\
\hline
\end{tabular}
\caption{The two possible states of the world (the null be either true or false) and the two possible decisions we can take given our data.}\label{tab:type12power}
\end{table}

\begin{figure}[!htbp]
\begin{knitrout}
\definecolor{shadecolor}{rgb}{0.969, 0.969, 0.969}\color{fgcolor}

{\centering \includegraphics[width=\maxwidth]{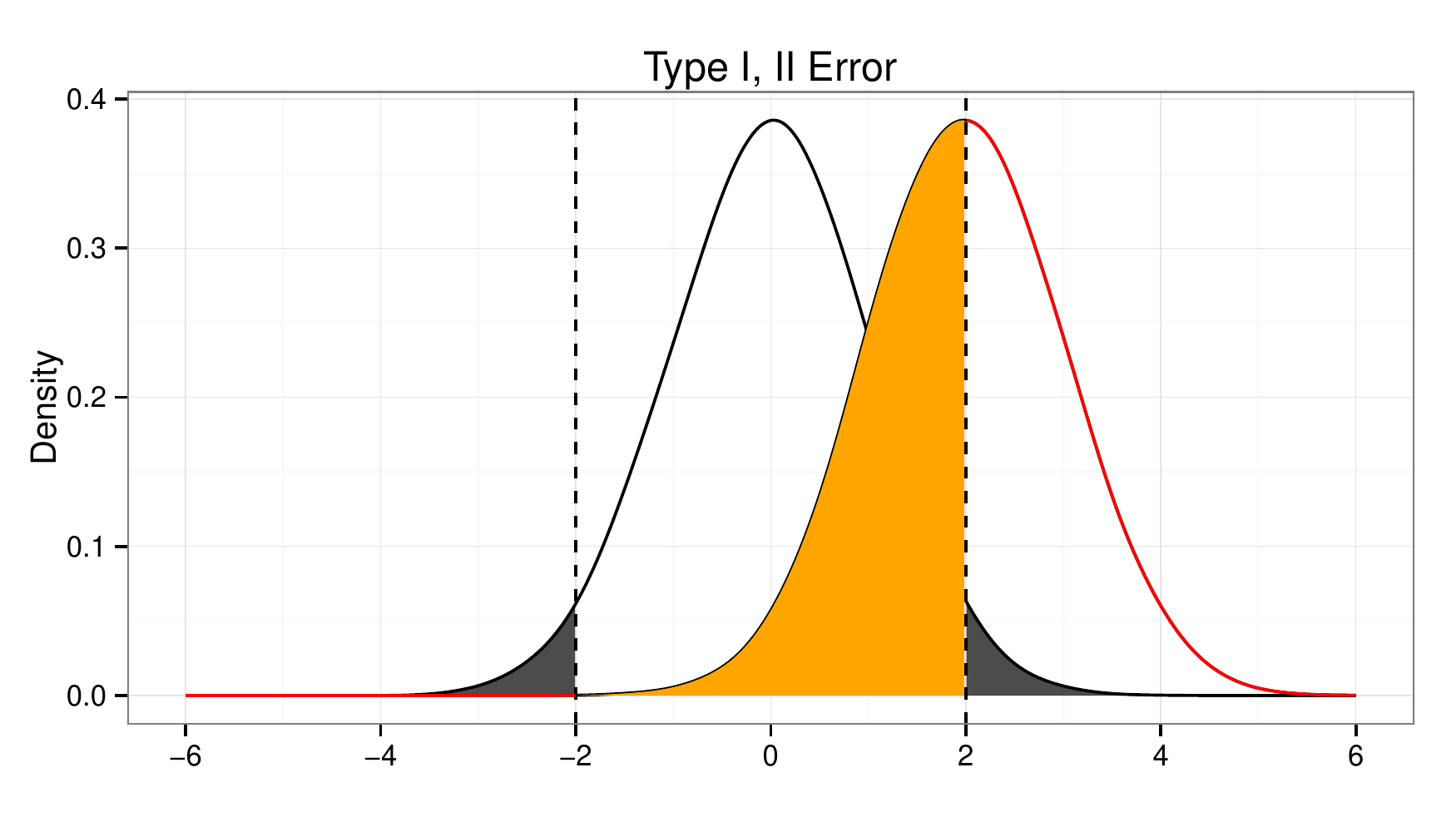} 

}

\end{knitrout}
\caption{An illustration of Type I error (the area shaded black) and Type II error (the area shaded orange) for a specific case where the true value of $\mu$ is 2.}\label{fig:type12}
\end{figure}

It may also help to see a visualization of Type I and II errors. Consider two different situations: in the first one, the true $\mu=0$; i.e., the null hypothesis is true; in the second one, the true $\mu=2$; i.e., the null hypothesis is false with a specific value for the alternative. Type I error will be relevant for the first situation, and it is illustrated as the black-colored area under the distribution representing the null hypothesis in Figure~\ref{fig:type12}. The area under the curve in these regions gives us the total probability of landing in these regions under the null. 
The figure also shows Type II error; this is the orange-colored region under the \textit{specific} alternative hypothesis $\mu = 2$.  We determine this area by first drawing the two vertical lines representing the points beyond which we would reject the null; then we compute the probability of landing within these points \textit{under the specific alternative}; this is exactly the orange-colored area. Power is not shown, but since power is 1-Type II error, it is all the area under the curve for the distribution centered around $2$ excluding the orange-colored area.

\begin{figure}[!htbp]
\begin{knitrout}
\definecolor{shadecolor}{rgb}{0.969, 0.969, 0.969}\color{fgcolor}

{\centering \includegraphics[width=\maxwidth]{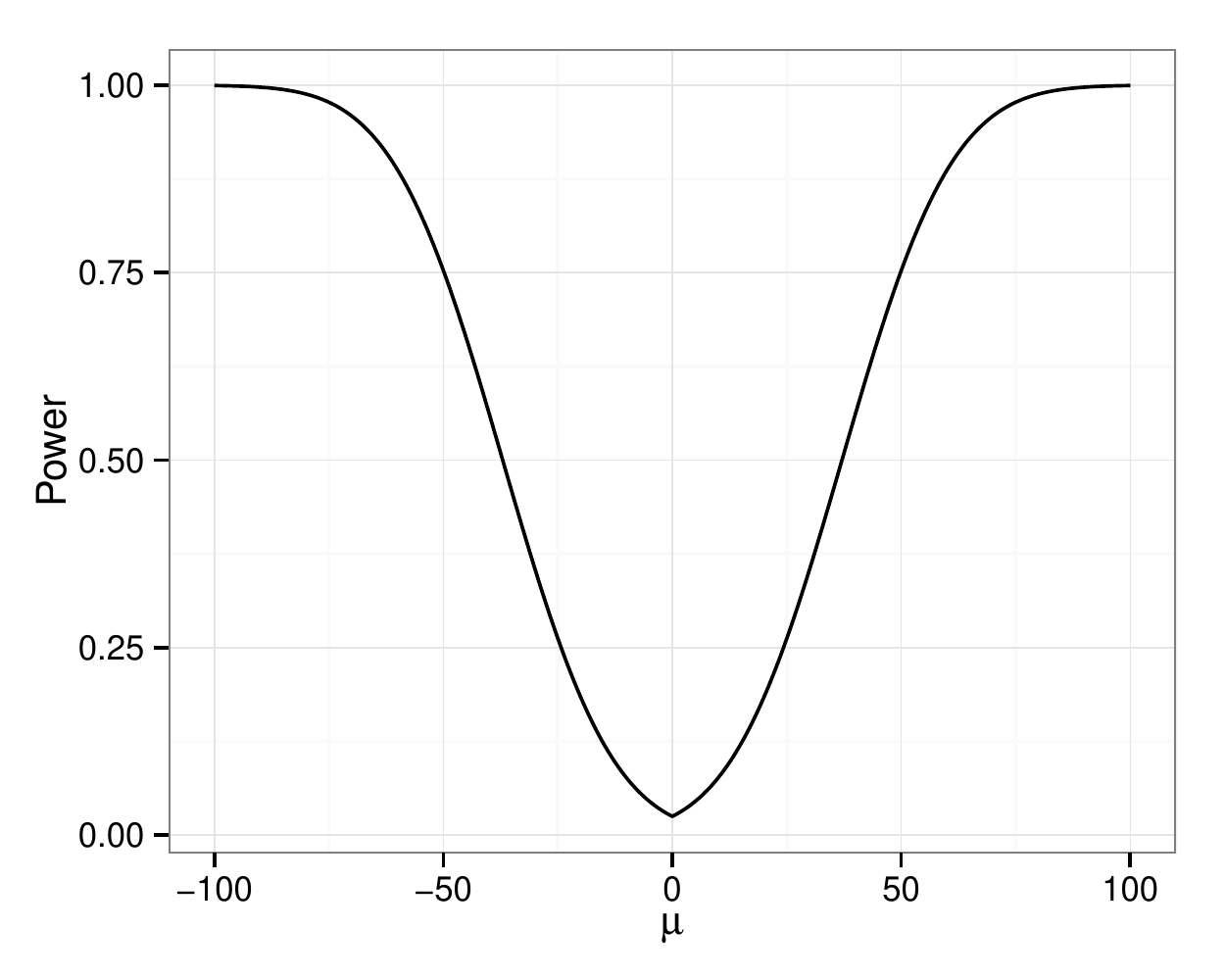} 

}

\end{knitrout}
\caption{An example of a power function for different effect sizes, assuming
(for illustration) a standard deviation of 40 and a sample size of
10.}\label{fig:powercurve}
\end{figure}

\begin{figure}[!htbp]
\begin{knitrout}
\definecolor{shadecolor}{rgb}{0.969, 0.969, 0.969}\color{fgcolor}

{\centering \includegraphics[width=\maxwidth]{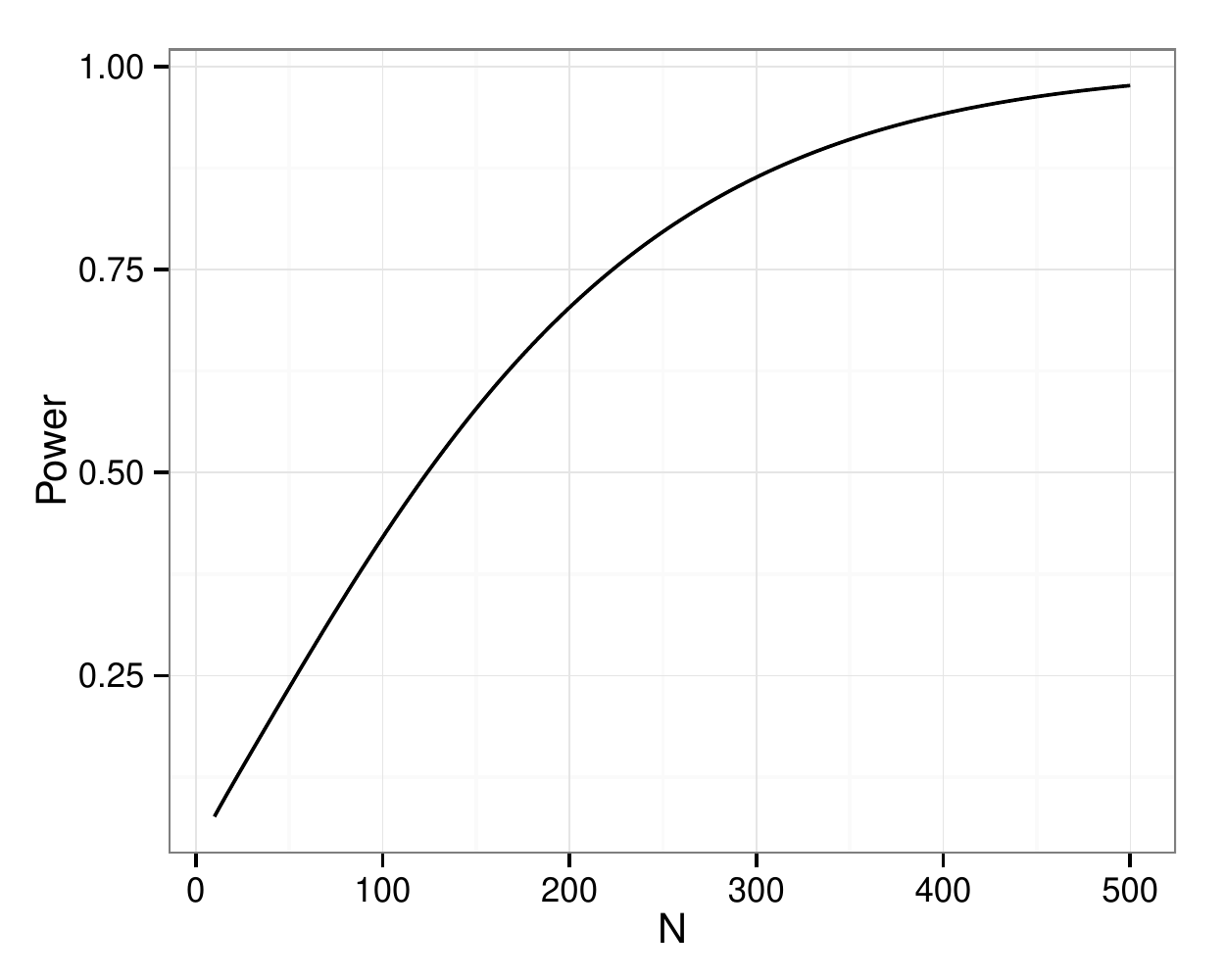} 

}

\end{knitrout}
\caption{An example of a power function for sample sizes (N), assuming (for
illustration) a standard deviation of 40 and a true effect of 10.}\label{fig:powercurveN}
\end{figure}

Power is best thought of as a function because we can only talk of power with reference to a specific alternative value for the magnitude (different from zero) of an effect ($\mu$), and the sample size and standard deviation of the sample. For different magnitudes $\mu$ of the effect we are interested in, the power will differ: in the scenario shown in Figure \ref{fig:type12}, the further away $\mu$ is from 0, the larger the power. The reader can verify this by considering what would happen to the Type II error region in Figure~\ref{fig:type12} if $\mu$ were 4 instead of 2. Clearly, Type II error would go down, and therefore power would increase.
Considering different values of $\mu$ in this manner, the power function is a curve that shows how power changes as $\mu$ changes (Figure~\ref{fig:powercurve}).
A practical implication is that one way to increase power is to design an experiment with a stronger manipulation,  one which will lead to a larger effect; for example, \citet{grodner} found larger effects when the sentence structure of interest was embedded inside another clause (also see \citet{barteketal09} for a replication of the Grodner and Gibson results). 
Another way to increase power and decrease Type II error
in a study is by increasing the sample size. 
Figure~\ref{fig:powercurveN} shows
how power changes as sample size changes.
Yet another way is to measure the dependent measure more precisely, thereby obtaining more accurate estimates of the standard deviation. For example, eyetracking data is extremely noisy, which may lead to an overestimate of the standard deviation. More frequent recalibration, using better equipment and well-trained experimenters could yield better estimates.

It is especially important to make sure you have high power if you are interested in arguing for the null. The reason for this should be obvious at this point: low power implies high Type II error, which means that any failure (even repeated failures) to reject the null may just be due to the fact that the probability of accepting the null when the null is in fact false is very high. There are many situations when you might want to argue for the null (see, e.g., \citet{Phillips-etal-2009}); here, it is imperative to put in extra effort into achieving as much power as possible (see \citet{JaegerBenzRoeserDillonVasishth2015} for an example).  

In order to be able to compute a reasonable estimate of power for a future study involving a comparison of two conditions, it is helpful to have an estimate of the difference
between the conditions in ms or log ms. Now, this of course
depends on having some way to determine a realistic estimate of the true
effect size for a particular phenomenon. This could be done through a meta-analysis or literature review \citep{VasishthetalPLoSOne2013,EngelmannJaegerVasishth2015} or computational modeling \citep{lewisvasishth:cogsci05}, or knowledge elicited from experts on the topic you are studying \citep{Vasishth:MScStatistics}. If such an estimate of the effect can be computed, 
then one can and should also compute Type S and M errors
\citep{gelmancarlin}. These are defined as follows:

\begin{enumerate}
\item
Type S error: the probability that the sign of the effect is incorrect, given
that (a) the result is statistically significant, or (b) the result is
statistically non-significant.
\item 
Type M error: the expectation of the ratio of the absolute magnitude of the
effect to the hypothesized true effect size (conditional on whether the result
is significant or not). Gelman and Carlin also call this the exaggeration
ratio, which is perhaps more descriptive than ``Type M error''.
\end{enumerate}

To illustrate Type S and M errors, we simulate  data with similar
characteristics as the data from
\citet{gibsonwu}; their experiment had a simple two-condition design and was originally
run with 40 participants and 16 items. As an example, we assume a
\emph{true} effect size of 0.01 for the log-transformed dependent variable, that is a difference of $\approx$4 ms from
a grand mean (the mean reading time of all the data) of 550 ms. 
The magnitude of the effect may strike the reader as smaller than they'd expect in a psycholinguistic study; however, note that we do not know the true effect, and previously published studies may be giving us overestimates (due to Type S and M errors) if they have low power. The choice of a small magnitude of effect here is just to illustrate what happens when power is low.

%Even though
%the magnitude may seem extremely small, we'll see next that unless we have a
%high-powered replication of a study, we may be seeing  exaggerated estimates
%in published papers.

For the simulation, we generated 1000 data sets, which we fit using
linear mixed models with a log-transformed dependent variable (using the package \emph{lme4}; \citet{lme4new}). We used models with varying intercepts only for 
subjects and for items, in 
order to have the highest power, at the cost of anticonservativity
\citep{matuschek2015balancing}; the discussion here does not depend on whether 
maximal models in the sense of \citet{barr2013random} are fit or not. The 
simulation shows the best-case scenario, since we generated the dependent 
variable (reading times) with a  perfectly lognormal distribution and no 
outliers; \citet{Ratcliff1993} shows that outliers can reduce power. 
The proportion of models under repeated sampling that show a significant effect  (an absolute observed t-value greater than two) gives us the power, which is only 0.09.  Type S error rate is 0.06 (calculated as the proportion of models with a significant effect, but with an estimated effect in the opposite direction of the true effect), and a Type M
error of 5.05 (calculated as the mean ratio of the
estimated effect to the true effect). 
This means that, in the present scenario, with a power of 0.09, 
we are likely to get an inflated estimate, and we have a  
6 percent probability of having the wrong sign if
we were to run the experiment repeatedly.

\begin{figure}[!htbp]
\begin{knitrout}
\definecolor{shadecolor}{rgb}{0.969, 0.969, 0.969}\color{fgcolor}

{\centering \includegraphics[width=\maxwidth]{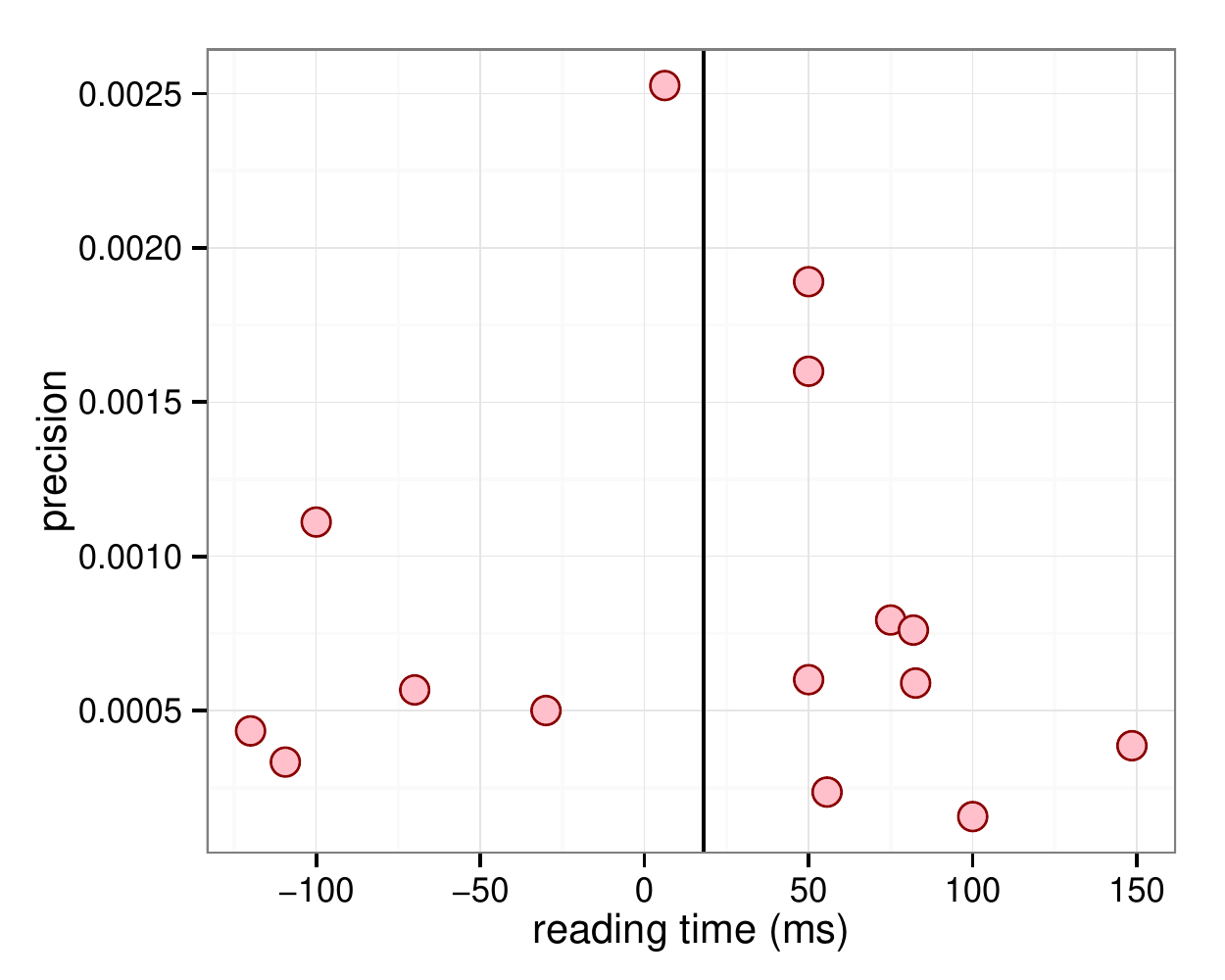} 

}

\end{knitrout}
\caption{A funnel plot for 15 Chinese relative clause studies data, showing precision plotted against mean effects. The vertical line marks the mean effect of $+18$ msec (which represents a subject relative advantage).}\label{fig:funnel}
\end{figure}

Thus, in low power situations, Type S and M error will be high; the practical consequence is that, if power is low, the magnitude and sign of the effect we find in our experiments may not be useful for calculating power in future experiments, for meta-analysis, or for modeling. This can be illustrated by considering the case of the Chinese relative clause controversy. There is a great deal of empirical disagreement regarding the question: in Chinese relative clauses, are subject relatives easier to process than object relatives (subject relative advantage)? Some researchers have claimed that object relatives are easier to process (at the head noun), and others have found the opposite, even in studies with the same design.
As discussed in \citet{Vasishth:MScStatistics}, we can see the Type S and M error in action in this real-life example. We use the funnel plot \citep{duval2000trim} for this purpose. When we display the means of 15 published studies, we see that studies with lower precision (the inverse of the variance, $1/SE^2$) have a bigger spread about the grand mean of these means (Figure~\ref{fig:funnel}). Since extreme values will be further away from the mean and will influence the grand mean, an obvious consequence is that if we want accurate estimates of the true effect, we should run as high-powered experiments as we can.\footnote{As an aside, note that such funnel plots can also be used to identify publication bias: gaps in the funnel are suggestive of unpublished data that should have been observed under repeated sampling.} 

Of course, in many situations (e.g., when studying special populations such as individuals with aphasia), there are practical limits to the sample size. Indeed, one could also argue that it is unethical to run unnecessarily large-sample studies because this would be a waste of resources and maybe even an abuse of participants. In such restrictive contexts, the researcher may be better off using a Bayesian framework with (mildly) informative priors (see part 2 of this review).  Our point here is that there is no substitute for attempting to calculate power before running an experiment, using the best estimates one can obtain. Note also that it is a mistake to use `observed' power, computed after the experiment has been run, to justify that one had adequate power for a study. At the very least, observed power could lead one astray if the ``true'' power happens to be actually quite low; in such a case, Type S and M errors will lead to overestimates of power.  Consider, for example, the fact evident from Figure~\ref{fig:funnel}, that a over-optimistically higher power will be estimated when using a low-precision study's estimates.

The next simulation illustrates the problem of low power by showing
potential differences between estimates and various true effect sizes. We
simulate the data of experiments similar to the one presented before but with
different values of true effect sizes (0.01, 0.02, 0.03, 0.05, 0.1) and in two flavors:
a small sample experiment (but still publishable) with 30
subjects and 16 items, and a medium-sized experiment with
80 subjects and 40 items.

\begin{figure}[!htbp]
\includegraphics{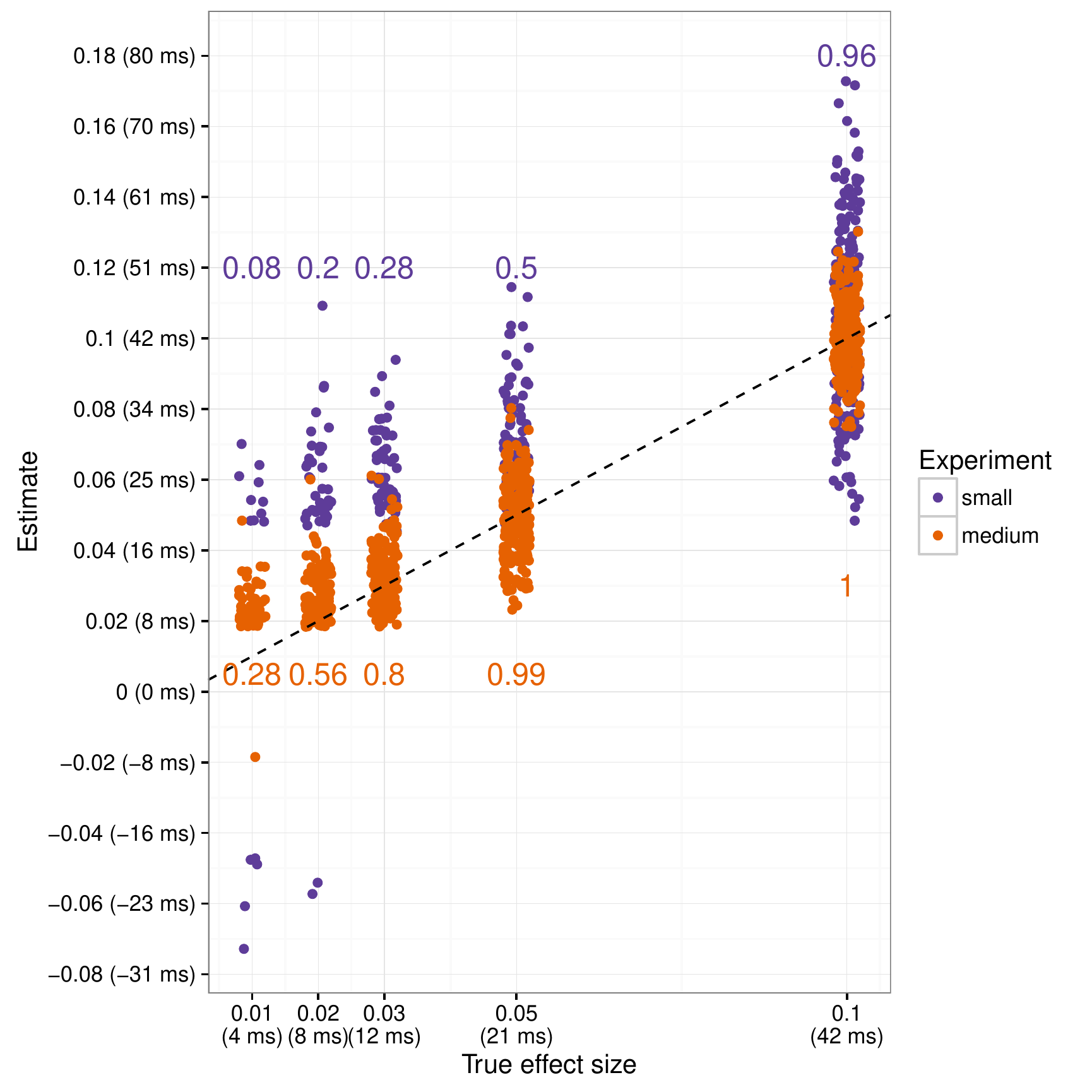} 
\caption{Plot illustrating the distribution of estimates under different true
effect sizes and two experiment sizes. Each point represents an experiment
with a significant effect, and for each effect size and experiment size, we
simulated 200 experiments. The power is shown within the figure for
the two sample sizes and for the different effect sizes.}\label{fig:estimates}
\end{figure}

Figure \ref{fig:estimates} shows the results of this simulation. We simulated
a total of ten (one small and one medium size experiment for each of the five
effect sizes) sets of 200 experiments. Every point in the graph
represents an experiment with a significant effect. The x-axis shows the true
effect size on the log scale, while the y-axis shows the estimate of the effect from linear mixed models with significant results. The power of each simulation also appears in the figure. The simulation shows that exaggerated
estimates (Type M errors) are more common for low-powered experiments. In
addition, when the underlying effect is very small, some experiments will show
results with the incorrect sign (Type S error). The dashed line shows the
ideal situation where the estimate and the effects are the same.

This means that if power is very low, and if journals are publishing
larger-than-true effect sizes---and we know that they have an incentive to do
so because researchers mistakenly believe that lower p-values, i.e.,
bigger absolute t-values, give stronger evidence for the specific alternative
hypothesis of interest---then power calculations based on published data are 
probably overestimating power, and thus also overestimating the replicability 
of our results.

%BN: is "To make matters worst" may be too strong?
%SV: you mean apart from being ungrammatical?
To make matters worse, assuming a normal distribution when the distribution is 
skewed and may have outliers can also reduce power \citep{Ratcliff1993}, thereby 
increasing Type S and M errors. Latencies
such as reading or response times are limited on the left by some amount of
time and they are right-skewed; as a consequence, assuming, as is standardly 
done in linguistics and psychology, that the underlying distribution generating
the data is normal can lead to loss of power.
Reading time or reaction time distributions are best fit with three
parameter distributions such as ex-Gaussian (the convolution of a Gaussian and
an exponential distribution), shifted lognormal (the log-transformed normal
distribution shifted to the right) or shifted Wald (the reciprocal of the
normal distribution shifted to the right); see, for example,
\citet{Luce1986,Ratcliff1993,Rouder2005}. These distributions, however, are
difficult to fit using standard frequentist software such the \texttt{lme4} 
function.  A middle-ground solution is to
apply a transformation on the dependent variable. This can reduce the impact
of  the skew (and of outliers) by  compressing larger values to a greater
extent than smaller values. The Box-Cox procedure
\citep{box1964analysis,Osborne2010} can be used to find out the best
transformation; but, for reading times, we find that the reciprocal-
(especially for self-paced reading) or the log-transformation (especially for
eye-tracking measures) are adequate, 
and have the advantage of being easy to interpret.
In sum, model assumptions matter. 
Even in the simple case of the one-sample t-test, violating the normality 
assumption leads to a fall in power.

A major practical implication of the Type M and S error problem is that if a 
study has low power, then it doesn't matter much whether you got a significant 
result  or not. Theory development based on low power studies would have a very 
weak empirical basis, regardless of the p-values obtained.
The main take-away point here is that we should run high powered studies, and attempt to replicate 
our results. There's really nothing as convincing as a replication. In our 
experience, reviewers and editors don't appreciate the importance of 
replications, but hopefully with increasing awareness of the issues, the culture will change.

\section{Running till significance is reached}

Given the importance that is attributed to significant findings regardless of
sample size, replicability, multiple comparisons, and so forth, an interesting
and troubling practice that we have encountered in linguistics and
psycholinguistics is running an experiment until significance is reached. The
experimenter gathers $n$ data points, then checks for significance (whether
$p<0.05$ or not). If the result is not significant, he/she gets more data ($n$
more data points) and checks for significance again. We can simulate data to
get a feeling for the consequences of such a procedure. A typical
initial $n$ might be $15$. This approach would give us a range of p-values
under hypothetical repeated sampling. Theoretically, under the standard
assumptions of frequentist methods, we expect a Type I error to be 
$0.05$.%This is the case in standard analyses.

%This gives us Type I error rate at about 5\%, as expected.

This is true if we fix the sample size in advance and repeatedly sample. 
But the statistical theory no longer holds if we adopt the strategy we
outlined above: run some participants, test for significance, and if
non-significant, run more participants. If we track the distribution of the
t-statistic for this approach, we will find that Type I error is much higher 
(in our simulation, approximately 15\%).

\begin{figure}[!htbp]
\begin{knitrout}
\definecolor{shadecolor}{rgb}{0.969, 0.969, 0.969}\color{fgcolor}

{\centering \includegraphics[width=\maxwidth]{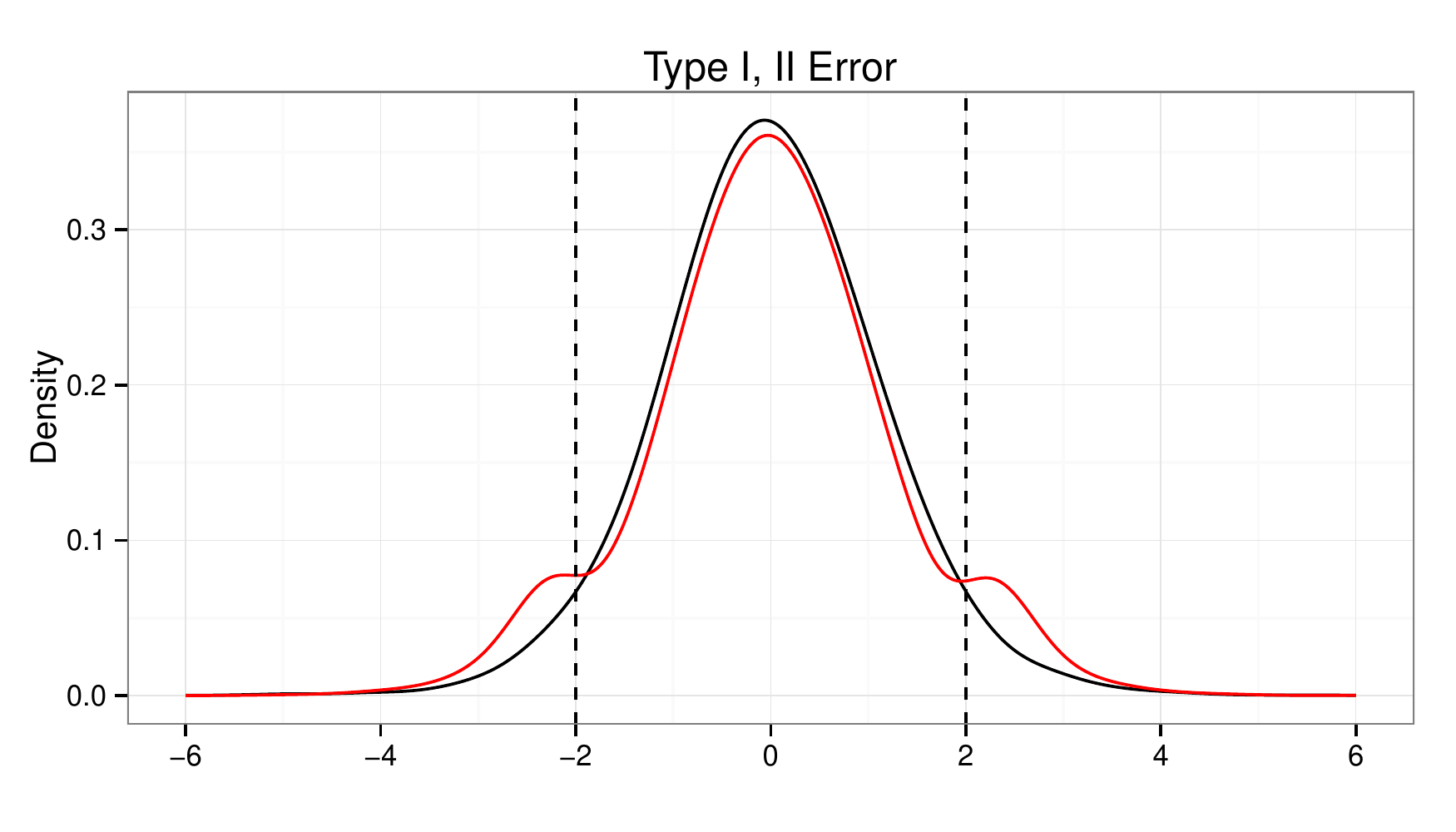} 

}

\end{knitrout}
\caption{The distribution of observed t-values under repeated sampling using
the stopping rule of run-till-significance. The dashed vertical lines mark the boundaries beyond which the p-value would be below 0.05.}\label{fig:stoppingrules}
\end{figure}

Let's compare the distribution, under repeated sampling, of the t-statistic in
the standard case vs with the above stopping rule (red) 
(Figure~\ref{fig:stoppingrules}).

We get bumps in the tails with the above stopping rule because,  under 
repeated sampling, some proportion of trials which have $p>0.05$ will be
replaced by trials in which $p<0.05$, leading to a redistribution of the 
probability mass in the t-distribution. This redistribution happens because 
we give ourselves more opportunities to get the desired $p<0.05$ under repeated 
sampling.
In other words, we have a higher Type I error than 0.05. It would of course 
be reasonable to take this approach if we appropriately adjust the Type I error; 
but this is not standard practice.
Thus, when using the standard frequentist theory,  one should fix one's sample 
size in advance based on a power analysis, not deploy a stopping rule like the 
one above; if we used such a stopping rule, we are much more likely to 
incorrectly declare a result as statistically significant.
Of course, if your goal is only to get a significant result so that you can 
get your article published, such stopping rules will give better results than 
fixing your sample size in advance!

\section{Multiple measures, multiple regions, degrees of freedom in analysis}

\citet{GelmanLoken2013} point out that there are in general too many ways to
analyze the data: from the choice of the statistical test to decisions on what
data points to exclude or include. This means that once we start looking hard
enough, it is possible to find a significant difference and then tell a
post-hoc theory that fits very neatly together. For that to happen, it is not
necessary that a researcher would go on a ``fishing expedition''
\citep{Gelman2013Slate}, that is, it is not necessary that he/she would be
actively trying to report any comparisons that happen to yield a
\emph{p}-value lower than 0.05; in many cases, the researcher just has too
many degrees of freedom in the analysis and it is not clear which is the right
way to analyze the data. 
A common example is to decide to report the result of a linear mixed model or an ANOVA or t-test, depending on which one of these yields a p-value below $0.05$. Researchers often flexibly switch between linear mixed models and repeated measures ANOVA to tell ``the best story'' they can given the data. One problem here is that using the ANOVA or t-test where a linear mixed model with crossed subject and item random effects is suggested by the design artificially reduces the sources of variance (through aggregation), with the result that effects that are not really statistically significant under a linear mixed model end up being significant once one aggregates the data. But the other problem with shopping around for the test that gives us the lowest p-value is the one that Gelman and colleagues point out:  we are introducing a degree of freedom in the analysis.

Another commonly seen situation is flexibly analyzing different regions of interest (often aggregating them post-hoc) until a $p<0.05$ result is found; for example, in \citet{Badecker-Straub-2002}, in their results section for experiment 5, they write: ``No significant differences emerged for any individual words or two-word regions. However when reading times are collapsed across the four positions following the reciprocal, reading times for this region were 48 ms longer in the multiple-match than in the single-match condition\dots''. This is an example of failing to find an effect in the region where it was expected a priori, and then trying to expand the regions of interest post-hoc.
Similarly, even when studying the same research question, researchers will sometimes trim the data, and sometimes not.

In sum, if the procedures have not been decided in
advance, we are no longer doing hypothesis testing, but an exploratory analysis \citep{deGroot2014}. Of course, there is no harm in doing and reporting exploratory analysis; but these analyses should be clearly marked as such and presented separately from the primary analyses.

Another issue is multiple comparisons.
In many experiments that use methods like self-paced reading, eye-tracking, and EEG, participants are presented with a whole sentence but
effects are expected to happen at a certain region. It is often not clear,
however, if the effect should appear at  only one specific word or several,
or even what the critical region is.
Furthermore, it is commonly assumed that effects can be delayed and appear in spillover
regions.  However, even though this is rarely acknowledged, fitting a model
(t-test, ANOVA, linear mixed model, etc.) for each possible word where the
effect may appear raises the problem of multiple comparisons, increasing the
chances of finding a false positive \citep{GelmanLoken2013}. In addition to the
multiple regions problem, eye-tracking-while-reading raises the problem that
multiple measures which are highly correlated (first fixation duration, single fixation duration, gaze time, etc.) are routinely analyzed as if they were separate sources of information \citep{vonderMalsburgAngele2015}. The motivation for analyzing multiples measures is that the theories being tested usually do not make explicit predictions about which measure is relevant. For EEG, even though it is not always the case, without a clear prediction about the ERP component, this issue is even more serious:  it is possible to look for effects in too many different groups of electrodes (which are correlated) and in different time windows \citep{FrankEtAlBrainLang2015}.

It is clear that in most cases researchers are not  doing tests until they
find a comparison showing a significant difference. In many novel experiments,
it is just not possible to know ahead of time where an effect will appear.
However, the problem of multiple comparisons may hinder replicability and give
a false feeling of certainty regarding the results.

We present three possible solutions to these problems. Linear mixed models can
solve the multiple comparisons problem if all relevant research questions can
be represented as parameters in one coherent hierarchical model
\citep{GelmanEtAl2012}, since the point estimates and their corresponding
intervals are shifted toward each other via ``shrinkage'' or ``partial
pooling'' \citep{GelmanEtAl2012}. However, building a single hierarchical
model that addresses all the research questions is not always trivial. For
several regions of interest, it may be possible to fit a single model using
Helmert contrasts (as in \citet{NicenboimEtAlFrontiers2015Capacity}). This
type of contrast compares each region with the average of the previous ones,
such that it is possible to discover a change in the pattern of the effects.
However, it is unclear if the effects should appear for all the trials in the
same region, since some participants in some trials could start a certain
process sooner predicting the structure of the item or could delay it due to
fatigue, lapse of attention, or because they have not finished a previous
cognitive process. Linear mixed models that can address the multiple measures
in eye-tracking or the highly multidimensional data of EEG are even more 
difficult to specify.

Another solution proposed recently by \citet{vonderMalsburgAngele2015} is to
fit independent models but to apply some type of correction such as
the Bonferroni correction. When new data can be easily gathered, a third 
possible solution is
to take results as exploratory until being confirmed with new data
\citep{deGroot2014,tukey}. The exploratory data is used to identify the
relevant regions, measures and/or ERP components, and only these potential
effects are then tested on the confirmatory analysis.  Researchers could pair
each new experiment with a preregistered replication
\citep{Nosek2012}, or gather more data than usual so that the full data set
could be divided into two subsets (for an example with EEG data, see
\citet{FrankEtAlBrainLang2015}).

\section{Conclusion}

In this article, we attempted to spell out some of the more common pitfalls associated with using the frequentist data-analytic approach.  The frequentist approach should be used as intended.   Given the growing realization that many claims in the scientific literature are false \citep{Ioannidis2005}, and the recent dramatic failure to replicate many of the published results \citep{Open2015}, now is as good a time as any to take stock and carefully consider how we use these tools. Checking model assumptions, running well-powered studies, avoiding exploratory analyses under the guise of hypothesis testing, and always replicating one's results; these are steps that are relatively easy to realize. In situations where data is sparse, Bayesian data analysis methods should be considered \citep{Gelman14}.

\section*{Acknowledgements}

We would like to thank Daniela Mertzen and Dario Paape for helpful comments on a draft.

\bibliography{LLC}

\end{document}